\documentclass[11pt,reqno]{amsart}

\pdfoutput = 1

\usepackage[english]{babel}
\usepackage[autostyle]{csquotes}
\usepackage{amsmath,amssymb,amsthm}
\usepackage{amsfonts}
\usepackage{amsxtra}

\usepackage{array,dcolumn}
\usepackage{graphicx}
\usepackage[hyperfootnotes=true,
        pdffitwindow=true,
        plainpages=false,
        pdfpagelabels=true,
        pdfpagemode=UseOutlines,
        pdfpagelayout=SinglePage,
                hyperindex,]{hyperref}

\usepackage{lmodern}
\usepackage[T1]{fontenc}
\usepackage[utf8]{inputenc}
\usepackage[activate={true,nocompatibility},spacing,kerning]{microtype}
\usepackage{bm}
\usepackage{bbm}
\usepackage[sc,osf]{mathpazo}
\microtypecontext{spacing=nonfrench}

 \calclayout

\newcommand{\mathsym}[1]{{}}
\newcommand{\unicode}[1]{{}}

\theoremstyle{plain}

\theoremstyle{definition}

\theoremstyle{remark}

\newcommand{\MeijerG}[8][\bigg]{G^{{ #2 },{ #3 }}_{{ #4 },{ #5 }} #1( \begin{matrix} #6 \\ #7 \end{matrix}\, #1\vert\, #8 #1)}

\numberwithin{equation}{section}

\begin{document}


\title{Meet Andr\'eief, Bordeaux 1886, and Andreev, Kharkov 1882--83}
\author{Peter J. Forrester}
\address{School of Mathematics and Statistics, 
ARC Centre of Excellence for Mathematical \& Statistical Frontiers,
University of Melbourne, Victoria 3010, Australia}
\email{pjforr@unimelb.edu.au}

\date{\today}


\begin{abstract}
The paper \enquote{Note sur une relation entre les int\'egrales d\'efinies des
produits des fonctions} by C.~Andr\'eief is
an often cited paper in random matrix theory, due to it
containing what is now referred to as Andr\'eief's integration formula.
 Nearly all citing works state
the publication year as 1883. However the journal containing the paper,
{\it M\'emories de la Societ\'e des Sciences physiques et naturelles de Bordeaux},
issue 3 volume 2 actually appeared in 1886. In addition to clarifying this point,
some historical information relating to C.~Andr\'eief (better known as K.A.~Andreev)
and the lead up to this work is given, as is a review of some  of the context of Andr\'eief's integration formula.
\end{abstract}


\maketitle


\section{Introduction}\label{s1}

Let $\{f_j(x) \}_{j=0}^{N-1}$ and $\{\phi_j(x) \}_{j=0}^{N-1}$ be two sequences of integrable functions.
Often encountered in random matrix theory is an eigenvalue probability density function (PDF)
proportional to the functional form
\begin{equation}\label{1}
\det [ f_j(x_k) ]_{j,k=0}^{N-1}  \det [ \phi_j(x_k) ]_{j,k=0}^{N-1}.
\end{equation}
The terminology used --- for reasons to be discussed in Section \ref{S3.2} below ---is that (\ref{1}) specifies a
biorthogonal ensemble \cite{Bor99}.

As some examples, choosing
\begin{equation}\label{fp}
f_j(x) = w(x) x^j, \qquad \phi_j(x) = x^j
\end{equation}
reduces (\ref{1}), upon applying the Vandermonde determinant identity (see e.g.~\cite[eq.~(1.173)]{Fo10}),
to
\begin{equation}\label{2}
\prod_{l=1}^N w(x_l) \prod_{1 \le j < k \le N} (x_k - x_j)^2.
\end{equation}
This is the eigenvalue PDF for so called unitary invariant ensembles of complex Hermitian matrices $X$.
More explicitly, with $w(x) = e^{\sum_{l=1}^\infty c_l x^l}$, the  matrices $X$ are to be chosen according to
the PDF proportional to $e^{\sum_{l=1}^\infty c_l {\rm Tr} X^l}$ --- this is unitary invariant in the sense that it
is unchanged by the mapping $X \mapsto U^\dagger X U$ for $U$ unitary.

As a generalisation of (\ref{fp}), but now with $x>0$, the choice
$$
f_j(x) = w(x) x^j, \qquad \phi_j(x) = x^{\theta j},
$$
again after use of the Vandermonde determinant identity, gives
\begin{equation}\label{2}
\prod_{l=1}^N w(x_l) \prod_{1 \le j < k \le N} (x_k - x_j) (x_k^\theta - x_j^\theta) .
\end{equation}
This class of eigenvalue PDF is said to define a Muttalib--Borodin ensemble; for realisations in terms
of random matrices involving complex Gaussian entries see
\cite{Ch14} and \cite{FW15}.

Let $Y = X_m X_{m-1} \cdots X_1$, where each $X_k$ be a complex standard Gaussian random matrix of size $N_k \times N_{k-1}$ with
$N=N_0 \le N_1 \le \cdots \le N_m$. The PDF of the squared singular values of $Y$ is of the form
(\ref{1}) with \cite{AIK13}
\begin{equation}\label{3}
f_j(x) = x^j, \qquad \phi_j(x) = 
 \MeijerG{m}{0}{0}{m}{-}{\nu_1,\ldots, \nu_{m-1},\nu_m + j}{x}.
\end{equation}
Here $G_{0,m}^{m,0}$ denotes a particular Meijer $G$-function and $\nu_k = N_k - N_0$. Also, as in
(\ref{2}), $x>0$. Modifications of the particular Meijer G-function in (\ref{3})
gives the PDF for the squared singular values of
products of mixtures of complex standard Gaussian random matrices and their inverses
\cite{Fo14}, and of truncations of random unitary matrices \cite{KKS15}. There is also a family
of matrix ensembles giving rise to PDFs of this type with $x$ taking values on all the real line
\cite{FIL18}. Another example with this latter property is given by the eigenvalue PDF for
Hermitian random matrices $(X - A)$ where $X$ is a member of the Gaussian unitary
ensemble (see e.g.~\cite[\S 1.3.1]{Fo10}) and $A$ is a fixed complex Hermitian matrix
with eigenvalues $\{ a_j \}$.
For this (see e.g.~\cite[\S 11.6.4]{Fo10})
\begin{equation}\label{fx}
f_j(x) = x^j, \qquad \phi_j(x) = e^{-x^2 + 2a_jx}
\end{equation}
Generally, a specialisation of (\ref{1}) with $f_j(x) = x^j$ is referred to as a polynomial
ensemble \cite{KZ14}.

The focus of this note is an integration formula associated with (\ref{1}), known as Andr\'eief's integration
formula \cite{An86}
\begin{multline}\label{4}
\int_I dx_1 \cdots \int_I dx_N \,
\det [ f_{j-1}(x_k) ]_{j,k=1}^N \det [ \phi_{j-1}(x_k) ]_{j,k=1}^N \\
= N! \det \Big [ \int_I f_j(x) \phi_k(x) \, dx \Big ]_{j,k=0}^{N-1},
\end{multline}
where $I$ denotes the support of $\{ f_j(x) \}$ and $\{ \phi_j(x) \}$. It turns out that there is some
confusion in the literature surrounding the publication date of this identity. Our aim is to set the record
straight on this point, and also take the opportunity to review the broader literature relating to (\ref{4}).

\section{The referencing of Andr\'eief's integration
formula, its proof and associated history}

\subsection{Referencing}
The identity (\ref{4}) is reported in a nearly all text books and extended reviews on random matrix theory:
\cite[Eq.~(5.3.7)]{Me04}, \cite[Th.~2.26]{TV04}, \cite[pg.~29]{ER05}, \cite[Lemma 2.2.2]{Bl09},
\cite[Eq.~(3.3)]{DG09}, \cite[Lemma 3.2.3]{AGZ09}, \cite[Eq.~(5.170)]{Fo10}, \cite[Prop.~4.2.5]{PS11}, 
\cite[\S 6.2]{BDS16}, \cite[Eq.~(11.1)]{LNV18}. In \cite{TV04}, \cite{ER05}, \cite{Bl09}, \cite{DG09}, \cite{BDS16} and
\cite{LNV18} the corresponding citation is given to \cite{An86} but with the publication year as 1883 rather than 1886.
In \cite{Me04} and \cite{Fo10} the identity is stated without reference, while \cite{AGZ09} references the 2nd edition
of the text book \cite{Me04}.
The work \cite{PS11} references the text books by Courant and Hilbert \cite[Section II.10.10]{CH53} and
P\'olya and Szeg\"o \cite[Problem II.68]{PS76} . The derivation in the latter is to take a continuum limit of the
Cauchy--Binet determinant formula; see \S \ref{S2.3} below. In the former the subsection in question is titled \enquote{A theorem on Gram's
determinant}, and as in \cite{PS76} it is presented as a continuous analogue of the Cauchy--Binet formula. In a somewhat
confusing twist, after citing  \cite[Section II.10.10]{CH53} the authors of \cite{PS11} proceed to name
(\ref{4}) \enquote{Gram's} theorem. 

The most recent of these references is \cite{LNV18}. In this the (mini) chapter containing (\ref{4}) is
titled \enquote{Meet Andr\'eief}, a play on which we've adopted as the title of the present note.

The identity (\ref{4}) is used in a very influential paper relating to random matrix theory by de Bruijn
\cite{dB55}, relating to a Pfaffian generalisation; see \ref{S3} below. Again the citation is to \cite{An86}, but
with a publication year of 1883. Similarly, the historical treatise of Muir \cite{Mu23} cites  \cite{An86} in
relation to (\ref{4}), but with the publication date of 1883.

A cited reference search on the Web of Science of the author name \enquote{Andreief, C.} gives around 60 papers citing \cite{An86} with a publication year of 1883,
although 3, namely \cite{BW14}, \cite{BL14}, \cite{HO14}, give the year of publication as 1886. 
Another source that gives the year of publication as 1886 is the G\"osta Mittag-Leffler separate collection 
\cite{ML}; see the 7th entry in section II \enquote{General function theory} of the Part I
\enquote{Small boxes} index.
The recent paper of Rosengren
\cite{Ro18} on determinantal elliptic Selberg integrals has reason to cite \cite{An86}. The publication year of 1886 is used, and
one finds the footnote  \enquote{Almost all citations to this paper states the publication year as 1883. For several reasons I
believe it was published in 1886, but I have not year been able to verify it.}

Some (on-line) research soon reveals that the University of Michigan's library holds a now digitized copy of
series 3, volume 2 of {\it M\'emories de la Societ\'e des Sciences physiques et naturelles de Bordeaux}.
The University of Michigan is/ was the address of J.~Baik and D.~Wang, the authors of \cite{BW14} giving the
publication year as 1886. Subsequence correspondence with them revealed that D.~Wang had in fact an electronic
copy of Andr\'eief's paper in his files. J.~Baik kindly accessed the full volume electronically and sent me the
relevant digitized pages for my own records. Indeed our (\ref{4}), with $N \mapsto N+1$ is on page 1 of
the citation \cite{An86}, and publication date is 1886. There is, nonetheless, some association with the year
1883: the paper is `signed', at the very end, \enquote{Kharkof, le d\'ecembre 1883}. In relation to an affiliation with Kharkof
(see also \S \ref{S2.4} below),
one reads in the preamble of the journal issue for 1886 that Mm.~Andreeff, professeur 
\`a l'Universit\'e de Kharkof, is one of the \enquote{Membres correspondents de la Soci\'et\'e}.

\subsection{Proof}
The proof of (\ref{4}) is very simple. The fact that the determinants in the integrand on the LHS are
anti-symmetric implies that upon the expansion
$$
\det [f_j(x_k) ]_{j,k=0}^{N-1} = \sum_{P \in S_N}
\varepsilon(P) \prod_{j=1}^N f_{j-1}(x_{P(j)}),
$$
where $S_N$ denotes the set of all $N!$ permutations $P$ and $\varepsilon(P)$ is their corresponding signature,
and taking the sum outside the integral, each integral has the same value. Thus the LHS is equal to
\begin{equation}\label{A.5}
N! \int_I dx_1 \cdots  \int_I dx_N \, \prod_{j=1}^N f_{j-1}(x_j) \, \det [ \phi_{j-1}(x_k) ]_{j,k=1}^N.
\end{equation}
Next, make use of the fact that the determinant of a matrix is equal to the determinant of its transpose to write
$$
 \det [ \phi_{j-1}(x_k) ]_{j,k=1}^N =  \det [ \phi_{k-1}(x_j) ]_{j,k=1}^N.
 $$
 Multiplying each factor $f_{j-1}(x_j)$, $j=1,\dots,N$ in the integrand of (\ref{A.5}) into the $j$-th row of this latter determinant gives
 the form
  $$
 N!  \int_I dx_1 \cdots  \int_I dx_N \,  \det \Big [ f_{j-1}(x_j) \phi_{k-1}(x_j) \Big ]_{j,k=1}^N.
 $$
 Here the dependence on $x_j$ is entirely in row $j$, so the integrations can be done row-by-row to give the
 RHS of (\ref{4}).
 
 \subsection{Relationship to the Cauchy--Binet formula}\label{S2.3}
 Let $X$ and $Y$ be  $M \times N$ $(M \ge N)$ matrices. Let $X_K$ (similarly $Y_K$) denote the restriction of $X$
 (similarly $Y$) to the rows indexed by $K$.
 The Cauchy--Binet formula states (see e.g.~\cite{Ai56})
  \begin{equation}\label{8}
  \sum_{K \subset \{1,\dots,M \}, \, |K| = N}
  \det [X_K] \det [Y_K] =
  \det [X^T Y ].
  \end{equation}
  As remarked above, many texts, including P\'olya and G.~Szeg\"o \cite{PS76},
  and Courant and Hilbert \cite{CH53}, demonstrate that  (\ref{4}) is a continuous
  version of (\ref{8}).
  
  One call also view (\ref{8}) as a discretisation of (\ref{4}).
  To see this, make the replacement $f_j(x) \mapsto \mu(x) f_j(x)$, then set
  $\mu(x) = \sum_{l=1}^M \delta (x - y_l)$ with $\{y_j\} \subset I$.
  The integration formula (\ref{4}) then reads
 $$
 \sum_{l_1=1}^M \cdots   \sum_{l_N=1}^M
 \det [ f_{j-1}(y_{l_k}) ]_{j,k=1}^N \det [ \phi_{j-1}(y_{l_k}) ]_{j,k=1}^N 
= N! \det \Big [  \sum_{l =1}^M f_j(y_l) \phi_k(y_l)  \Big ]_{j,k=0}^{N-1},
$$
 or equivalently, upon ordering $\{l_j\}_{j=1}^N$,
  \begin{equation*}\label{9}
 \sum_{1 \le l_1 < l_2 < \cdots < l_N \le M}
 \det [ f_{j-1}(y_{l_k}) ]_{j,k=1}^N \det [ \phi_{j-1}(y_{l_k}) ]_{j,k=1}^N 
=  \det \Big [  \sum_{l =1}^M  f_j(y_l) \phi_k(y_l)  \Big ]_{j,k=0}^{N-1}.
\end{equation*}
 Defining the $M \times N$ matrices $X$ and $Y$ by
 $$
 X^T = [f_{j-1}(y_l) ]_{j=1,\dots,N \atop l = 1,\dots, M}, \qquad
 Y^T = [ \phi_{j-1}(y_l) ]_{j=1,\dots,N \atop l = 1,\dots, M},
 $$
 one recognises (\ref{9}) as (\ref{8}).

 \subsection{Some history}\label{S2.4}
 The paper \cite{An86} begins by stating that the motivation comes from the work of Chebyshev on a class
 of inequalities (sometimes referred to as Chebyshev's `other' inequality). As is made
 clear in the historical article by Mitrinovi\'c and Vasi\'c \cite{MV74}, one of the simplest examples
 in this class is the result
 \begin{equation}\label{7a}
 \int_a^b f(x) g(x) \, dx \ge {1 \over b - a} \int_a^b f(x) \, dx 
 \int_a^b g(y) \, dy,
 \end{equation}
 subject to both $f$ and $g$ being increasing, or both decreasing, on the interval $(a,b)$.
 
 Consider now (\ref{4}) in the case $N=2$ with $I = [a,b]$. Expanding the determinants gives
\begin{multline}\label{7a.1}
{1 \over 2} \int_a^b dx_1 \int_a^b dx_2 \,
\Big (f_0(x_1) f_1(x_2) - f_1(x_1) f_0(x_2) \Big )
\Big (\phi_0(x_1) \phi_1(x_2) - \phi_1(x_1) \phi_0(x_2) \Big )
\\
= \Big ( \int_a^b f_0(x) \phi_0(x) \, dx \Big )
 \Big ( \int_a^b f_1(x) \phi_1(x) \, dx \Big ) -
 \Big ( \int_a^b f_0(x) \phi_1(x) \, dx \Big )
 \Big ( \int_a^b f_1(x) \phi_0(x) \, dx \Big ) .
\end{multline} 
 Setting $f_0(x) = f(x)$, $f_2(x) = g(x)$, $\phi_0(x) = \phi_1(x) = 1$ this reads
\begin{multline*} 
(b - a) \int_a^b f(x) g(x) \, dx - \int_a^b f(x) \, dx  \int_a^b g(x) \, dx \\
= {1 \over 2} \int_a^b \int_a^b \Big (f(x) - f(y) \Big ) \Big (g(x) - g(y) \Big ) \, dx dy.
\end{multline*} 
By the assumption on $f$ and $g$ being always increasing or always decreasing on $(a,b)$, the
RHS is non-negative and thus so is the LHS; this latter fact is (\ref{7a}).

From \cite{MV74} one learns that (\ref{7a.1}) appeared in an earlier work of Andr\'eief
\cite{An82}, written in Russian. But now, using the Roman alphabet, the author's name is
written K.A.~Andreev rather than C.~Andr\'eief. In fact, as sourced by my colleague Kostya
Borovkov, this article is available online at
 http://kms.univer.kharkov.ua/authors.html; the English translation of the citation for \cite{An82},
 as made by K.~Borovkov, is given in the references.
 One learns that the argument given in \cite{MV74} leading from (\ref{7a.1}) to
(\ref{7a}) is that contained in  
 \cite{An82}, and furthermore that the publication of \cite{An82} was in the second issue of
 the 1882 volume, not 1883 as cited in \cite{MV74}.
 
 K.~Borovkov also pointed me too a well sourced and informative Wiki page on
 K.A.~Andreev \cite{WiAndreev} --- full name Konstantin Alekseevich Andreev (14 March 1848 -- 29 October 1921).
 Relevant to the `signature' at the end of the article \cite{An86} as referred to above, one learns that for around a 20 year period from
 1879 he was a full professor at the Kharkov University, as well as the Kharkov Technology Institute.
 Also, he has an association with France between the writing of his PhD and his habilitation, working in Paris
 (as well as Berlin).
 
 \section{de Bruijn's integration formula and orthogonality}\label{S3}
 \subsection{de Bruijn's integration formula}
 Let $h(x,y) = - h(y,x)$ and thus be antisymmetric so that ${\rm Pf} \, [h(x_j, x_k) ]_{j,k=1}^{2n}$ is well defined.
 As a generalisation of (\ref{4}), de Bruijn obtained the integration formula
\begin{multline}\label{4B}
{1 \over (2n)!} \int_I dx_1 \cdots \int_I dx_{2n} \,
\det [ f_{j-1}(x_k) ]_{j,k=1}^{2n} {\rm Pf} [ h(x_j,x_k) ]_{j,k=1}^{2n} \\
= {\rm Pf} \Big [ \int_I dx \int_I dy \, f_{j-1}(x) h(x,y) f_{k-1}(y) \Big ]_{j,k=1}^{2n},
\end{multline} 
 and also an analogous formula in the case that $\det [ f_{j-1}(x_k) ]$ is odd rather than even in size. This generalises
 (\ref{4}) in the sense that its discretisation includes the Cauchy--Binet formula as a special case.
 
 To see this, let $N \le M$ be positive integers, $T$ be an $N \times M$ matrix, and $A$ be an $M \times M$ anti-symmetric matrix.
 As noted in \cite{BR01a}
 the discretisation procedure used to deduce the Cauchy--Binet formula from (\ref{4}) then gives  the
 so called minor summation formula \cite{IOW96}
  \begin{equation}\label{AT}
  \sum_{J \subset \{1,\dots,M \}, \, |J| = N}
 {\rm Pf} [A_{J,J} ]  \det [T_J] =
  {\rm Pf} [T A^T T ].
  \end{equation}
  Here $A_{J,J}$ denotes the restriction of the rows and columns of $A$ to the set $J$, while $T_J$ denotes the restriction of
  the columns to that set. It is shown in \cite{BR01a} (see also \cite[Exercises 6.3 q.4]{Fo10}) that upon setting $M = 2m$
  and $N = 2n$, the choice
  $$
  A = \begin{bmatrix}  O_m & \mathbb I_m \\
  -  \mathbb I_m &  O_m \end{bmatrix}, \qquad
  T = \begin{bmatrix} X_{m \times n} & O_{m \times n} \\
  O_{m \times n} &  Y_{m \times n} \end{bmatrix},
$$ 
 reclaims (\ref{8}).
 \subsection{Relationship to orthogonal functions}\label{S3.2}
 From the Gram--Schmidt orthogonalisation procedure, associated with
 $\{f_j(x)\}_{j=0}^{N-1}$ and $\{\phi_j(x)\}_{j=0}^{N-1}$ in (\ref{1}) are biorthgonal functions 
 $\{F_j(x)\}_{j=0}^{N-1}$ are $\{\Phi_j(x)\}_{j=0}^{N-1}$ with the triangular structure
   \begin{equation}\label{T}
 F_j(x) = f_j(x) + \sum_{l=0}^{j-1} c_{jl} f_l(x), \qquad
 \Phi_j(x) = \phi_j(x) + \sum_{l=0}^{j-1} d_{jl} \phi_l(x),
   \end{equation}
 where $\{ c_{jk} \}$, $\{ d_{jk} \}$ scalars, and the biorthogonality relation
  \begin{equation}\label{T1} 
 \int_I F_j(x) \Phi_k(x) \, dx = h_j \delta_{j,k}
 \end{equation}
 for some normalisations $\{h_j\}$.
 The triangular structure (\ref{T}) implies that the determinants in (\ref{1}) are left unchanged upon
 use of the biorthogonal functions,
 $$
\det [ f_j(x_k) ]_{j,k=0}^{N-1}  = \det [ F_j(x_k) ]_{j,k=0}^{N-1}, \qquad
 \det [ \phi_j(x_k) ]_{j,k=0}^{N-1} =  \det [ \Phi_j(x_k) ]_{j,k=0}^{N-1}.
 $$
 This explains the use of the term biorthogonal ensemble used in relation to (\ref{1}) (in the special case
 that $f_j(x) = x^j$, each $F_j(x)$ is a polynomial of degree $j$, and this is the reason form the
 terminology polynomial ensemble noted below (\ref{fx})). Upon the use of the biorthogonal functions, (\ref{4})
 simplifies to 
 \begin{equation}\label{4S}
\int_I dx_1 \cdots \int_I dx_N \,
\det [ f_{j-1}(x_k) ]_{j,k=1}^N \det [ \phi_{j-1}(x_k) ]_{j,k=1}^N 
= N! \prod_{j=0}^{N-1} h_j,
\end{equation}
where the $h_j$ are in (\ref{T1}). A similar simplification of (\ref{4B}) is possible, upon the introduction of skew orthogonal
functions; see e.g.~\cite[Ch.~6]{Fo10} for some special cases.
 
   \section*{Acknowledgements}
The assistance of J.~Baik, K.~Borovkov and D.~Wang  in providing source material has been essential to the
writing of this note. 
The research project itself is part of the program of study supported by the 
ARC Centre of Excellence for Mathematical \& Statistical Frontiers,
and the Australian Research Council Discovery Project grant DP170102028.
 
\nopagebreak

\providecommand{\bysame}{\leavevmode\hbox to3em{\hrulefill}\thinspace}
\providecommand{\MR}{\relax\ifhmode\unskip\space\fi MR }
\providecommand{\MRhref}[2]{%
  \href{http://www.ams.org/mathscinet-getitem?mr=#1}{#2}
}
\providecommand{\href}[2]{#2}

\end{document}